\documentclass[useAMS,usenatbib]{mn2e}
\usepackage{graphicx}
\usepackage{rotating,times,pictex,graphicx,latexsym}
\usepackage{color}
\usepackage{longtable,amsmath}
\usepackage{lscape}

\title{UKIRT follow-up observations of the old open cluster FSR\,0358
(Kirkpatrick\,1)} 

\author[Froebrich, et al.]{D.~Froebrich$^{1}$\thanks{E-mail:
df@star.kent.ac.uk}, H.~Meusinger$^{2}$, C.J.~Davis$^{3}$ and S.~Schmeja$^{4}$\\
$^1$ Centre for Astrophysics and Planetary Science, University of Kent,
Canterbury, CT2 7NH, UK \\ $^2$ Th\"uringer Landessternwarte Tautenburg,
Sternwarte 5, 07778 Tautenburg, Germany \\ $^3$ Joint Astronomy Centre, 660
North A\`{}ohoku Place, University Park, Hilo, Hawaii 96720, USA \\ $^4$ Zentrum
f\"ur Astronomie der Universit\"at Heidelberg, Institut f\"ur Theoretische
Astrophysik, Albert-Ueberle-Str. 2, 69120 Heidelberg, Germany } 

\begin{document}

\date{Received sooner; accepted later}
\pagerange{\pageref{firstpage}--\pageref{lastpage}} \pubyear{2007}
\maketitle

\label{firstpage}

\begin{abstract}

We aim to characterise the properties of the stellar clusters in the Milky Way.
Utilising an expectation-maximisation method we determined that the cluster
FSR\,0358, originally discovered by J.D.\,Kirkpatrick, is the most likely real
cluster amongst the cluster candidates from Froebrich et al.. Here we present
new deep high resolution near infrared imaging of this object obtained with
UKIRT. The analysis of the data reveals that FSR\,0358 (Kirkpatrick\,1) is a
$5\pm2$\,Gyr old open cluster in the outer Milky Way. Its age, metallicity of
Z\,=\,0.008 and distance from the Galactic Centre of 11.2\,kpc are typical for
the known old open galactic clusters. So far six of the FSR cluster candidates
have been identified as having an age above 5\,Gyr. This shows the significance
of this catalogue in enhancing our knowledge of the oldest open clusters in the
Galaxy.

\end{abstract}


\section{Introduction}

Star clusters are fundamental building blocks of galaxies. Observations from our
local vicinity in the Galactic disk suggest that the majority of stars form in
embedded star clusters of which a fraction survives to become open or globular
clusters. The understanding of clustered star formation is closely related to
various issues of the evolution of the Galaxy itself (e.g. West et al.
\cite{2004Natur.427...31W}, Kroupa \cite{2008arXiv0810.4143K}). One of the
pre-requisites for corresponding studies is the availability of large and
homogeneous cluster samples which do not strongly suffer from major selection
effects.

The development of observing facilities in the infrared has provided not only
the ability to study in detail young star clusters embedded within molecular
clouds but also to search for clusters which are unseen in the optical due to
obscuration from dust. The all-sky surveys carried out during the last decade in
the near infrared (NIR) provided the possibility to improve the completeness of
the catalogues of known star clusters in the Galaxy significantly.  Recent
studies have shown that embedded infrared open clusters might outnumber the
cluster population detected in the optical by an order of magnitude (see Lada \&
Lada \cite{2003ARA&A..41...57L} for a review). But also old open clusters and
globular clusters located in sky regions of strong foreground extinction
($A_{\rm V} \approx 10$\,mag or above) have a significantly higher probability
to be discovered in the NIR than in the optical. As for the all-sky distribution
of the galaxies from optical surveys, the database of optically detected old
star clusters in our Galaxy is expected to be substantially incomplete in the
area close to the Galactic Plane. The 2 Micron All Sky Survey (2MASS - Skrutskie
et al. \cite{2006AJ....131.1163S}) provides an excellent database to identify a
large sample of infrared star clusters in this so-called {\it Zone of
Avoidance}, among them many which were previously unidentified.

We have carried out a systematic large-scale cluster search at low Galactic
latitudes ($|b| <$\,20\degr) based on star counts in 2MASS (Froebrich, Scholz \&
Raftery \cite{2007MNRAS.374..399F}; hereafter FSR). The FSR survey identified
more than 1000 previously unknown potential clusters, among them several
promising candidates for globular clusters. First results from the detailed
analysis of deeper NIR follow-up observations with higher spatial resolution for
the high-priority globular cluster candidates FSR\,1735 and FSR\,0190 confirmed
their classification as very old systems (Froebrich, Meusinger \& Scholz
\cite{2007MNRAS.377L..54F}; Froebrich, Meusinger \& Davis
\cite{2008MNRAS.383L..45F}). Other globular cluster candidates from the FSR list
were discussed by Bica et al. \cite{2007A&A...472..483B} and Bonatto et al.
\cite{2007MNRAS.381L..45B}. The analysis of deep follow-up observations with the
ESO/NTT for a sample of 14 FSR cluster candidates by Froebrich, Meusinger \&
Scholz \cite{2008MNRAS.390.1598F} revealed 7 real stellar clusters, in agreement
with the pre-estimated success rate of 50\,\% for the FSR cluster search, among
them several clusters with estimated isochrone ages of several Gyr. This paper
also gives an overview of all so far investigated FSR cluster candidates.

The present paper is concerned with another high-priority old cluster candidate,
FSR\,0358, which has a striking resemblance to FSR\,0190. Neither in the SIMBAD
nor in the WEBDA  database an entry was found for a star cluster near the
position of FSR\,0358 and no citation was found in refereed publications.
However, there is a quotation of this cluster on the 2MASS Atlas Image Gallery
webpage for open clusters\footnote{\tt
http://www.ipac.caltech.edu/2mass/gallery/images\_open.html} containing a
comment on its analysis by J.D.\,Kirkpatrick. Given that the cluster designation
should be that of the discoverer, the official name of the cluster is of course
Kirkpatrick\,1. Nevertheless we will use the designation FSR\,0358 throughout
this paper since it is part of a series of publications based upon deeper
follow-up observations of cluster candidates from the FSR catalogue.   

The analysis of the 2MASS data by Kirkpatrick using solar metallicity isochrones
from Girardi et al. \cite{2002A&A...391..195G} suggests that the cluster is at a
distance of 7.2\,kpc and that there are about 5\,mag optical extinction along
the line of sight. Here we will investigate this interesting object in more
detail. Compared with 2MASS our new observations allow a significantly deeper
photometry with substantially reduced image crowding problems due to improved
spatial resolution and are thus suited to check the cluster parameters derived
by Kirkpatrick. In particular, our new data has the potential to detect, for the
first time, the main sequence turn-off and to analyse the sub-giant branch to
make the age estimation more robust.

The paper is structured as follows. In Sect.\,\ref{data} we describe our
observations, data reduction and photometry. Section\,\ref{results} details our
analysis and discussion of the cluster based on our new data, supplemented by
2MASS. The conclusions are presented in Sect.\,\ref{discussion}.

\section{Data}
\label{data}

\subsection{Target Selection}

The FSR list of potential cluster candidates is judged to be contaminated with
about 50\,\% of random star density enhancements (Froebrich et al.
\cite{2007MNRAS.374..399F,2008MNRAS.390.1598F}). To identify the most probable
real star clusters in the sample, we have modelled the cluster candidates by
two-dimensional angular Gaussian distributions applying an
expectation-maximization algorithm (Dempster, Laird \& Rubin \cite{dempster77})
and evaluating the best fit using the Bayesian information criterion (BIC,
Schwarz \cite{schwarz78}). This technique has for example been used to identify
star clusters in the Glimpse data by Mercer et al. \cite{2005ApJ...635..560M}.
We find that FSR\,0358 has the lowest BIC value of all the FSR candidates
(Schmeja \& Froebrich in prep.), indicating the highest probability of being a
real cluster. Hence it is one of the most interesting clusters to investigate by
follow-up observations.

\subsection{Observations and Data Reduction}

FSR\,0358 was observed in the Summer of 2008 at the United Kingdom Infrared
Telescope (UKIRT) with the facility imager-spectrometer, UIST (Ramsay Howat et
al. \cite{2004SPIE.5492.1160R}). UIST contains a 1024x1024 InSb array; the plate
scale measures 0.12\arcsec\ per pixel. Mosaics of FSR\,0358 and an adjacent
control field were obtained through Mauna Kea Consortium J ($\lambda =
1.25$\,$\mu$m, $\delta\lambda = 0.16$\,$\mu$m), H ($\lambda = 1.64$\,$\mu$m,
$\delta\lambda = 0.29$\,$\mu$m) and K-band ($\lambda = 2.20$\,$\mu$m,
$\delta\lambda = 0.34$\,$\mu$m) filters.

To map an area covering 7.1\arcmin\,$\times$\,7.1\arcmin\ centred on FSR\,0358
(map centre: 22:10:12.8 +58:48:12 (J2000)), in each filter four 8-point mosaics
were obtained. Directly after observing FSR\,0358 at each wavelength, an
18-point mosaic of the control field was secured (map centre: 22:10:12.8
$+$58:38:12 (J2000); area covered $\sim$\,5.3\arcmin\,$\times$\,5.3\arcmin). All
J and K-band data were obtained on the 16$^{\rm th}$ of August; all H-band data
were observed on the 22$^{\rm nd}$ of August (note that UKIRT is fully flexibly
scheduled). The control field was observed at essentially the same airmass and
under the same observing conditions as FSR\,0358. All data were acquired under
photometric skies. The spatial resolution due to seeing, telescope and
instrumental effects, measured from Gaussian fitting of stars in the reduced
mosaics, was $\sim$0.65\arcsec\ in J and $\sim$\,0.70\arcsec\ in H and K.

All images were dark-subtracted and bad-pixel-masked before flat-fielding using
the median average of the frames themselves. For the control field all 18 frames
in each filter were reduced together. For FSR\,0358, the 8 images in the four
quadrants of each broad-band mosaic were trimmed and mosaicked together, before
these four corner mosaics were registered and combined to produce the final
7.1\arcmin\,$\times$\,7.1\arcmin\ mosaics. The four corner mosaics were
corrected for extinction before this last mosaicking step. Stars in the
overlapping regions were used for astrometric and photometric registration. 

\subsection{Photometry}

Our new observations are about 3\,mag deeper than 2MASS and of higher spatial
resolution. Despite this, and the fact that the cluster is just 2.2\degr\ above
the Galactic Plane, the crowding of the field is only of the order of 1.3\,\%
(see also Fig.\,\ref{kband}). We have hence performed the identification of
stars and their photometry using the Source Extractor software (Bertin \&
Arnouts \cite{1996A&AS..117..393B}). Since we are interested only in high
quality photometry in all three filters, the search for objects was performed in
the J-band mosaic. Objects were selected as real, only if they consisted of more
than 25 connected pixels (the seeing) with fluxes more than 2\,$\sigma$ above
the background noise. The subsequent photometry in the H- and K-band image was
then only performed on objects detected in the J-band. There were 5211 objects
detected in the cluster field and 1924 in the control field. 


The field of the cluster, as well as the control field, contains a large number
of stars, identifiable in 2MASS. These were used to calibrate our photometry
into the 2MASS system. The {\it rms} uncertainties in the calibration are 0.1,
0.4, and 0.25\,mag for J, H, and K in the control field, and 0.2, 0.15, and
0.17\,mag in the cluster field. This rather large scatter is caused by the much
better spatial resolution of our new images compared to 2MASS. Note that stars
brighter than about K\,=\,13\,mag are saturated in our new images.

Given the fact that our cluster field was observed in four different parts,
taken at slightly different air masses, we investigated if calibrating the
different parts of the mosaic individually will influence our results. We find
that the difference in the calibration is much smaller than the {\it rms}
scatter, and hence there is no significant influence onto our results.

\section{Results}
\label{results}

\subsection{Cluster Position and Appearance}

\begin{figure}
\centering
\includegraphics[angle=-90,width=8cm]{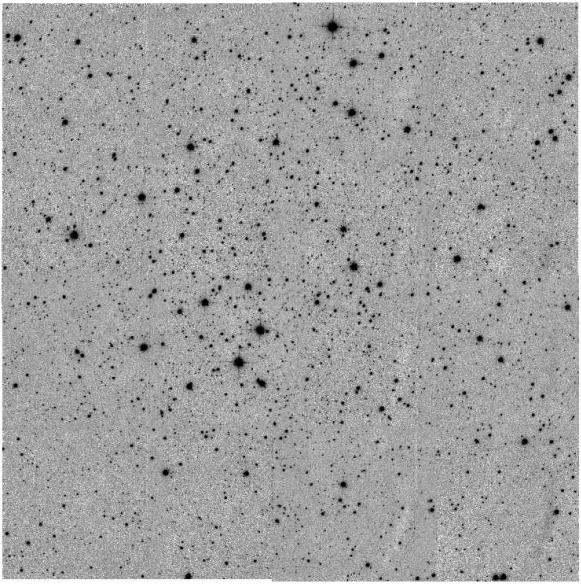}
\caption{\label{kband} Gray scale representation of the K-band image of
FSR\,0358. The image is 7.15\arcmin\ $\times$ 7.15\arcmin\ in size. The map
centre is at 22:10:12.8 +58:48:12 (J2000). North is to the top and East to the
left.} 
\end{figure}

The cluster image (for a gray scale representation see Fig.\,\ref{kband}) shows
an extended cluster of stars. The actual centre is not well defined. We find
that the highest star density in our new UKIRT data occurs at RA\,=\,22:10:14.1
and DEC\,=\,$+$58:48:02 (J2000). This corresponds to the Galactic Coordinates of
$l$\,=\,103.34094\degr\ and $b$\,=\,+2.20214\degr, about half an arcminute away
from the position originally listed in Froebrich et al.
\cite{2007MNRAS.374..399F}, as well as from the position given by Kirkpatrick.

\subsection{Star Density Profile}

\begin{figure}
\centering
\includegraphics[angle=-90,width=8cm]{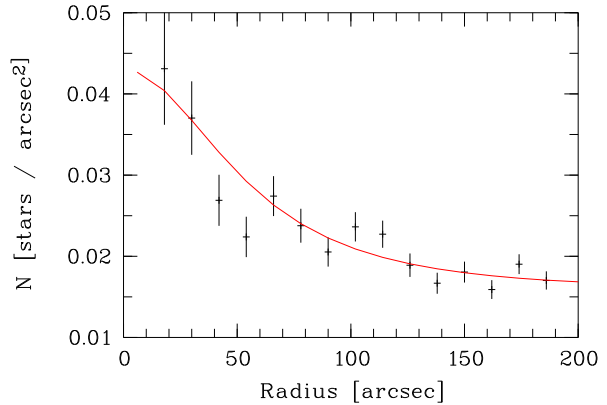}
\caption{\label{stardens} Star Density profile using the new UKIRT data. The
solid line represents a fit using a core radius of 60\arcsec.} 
\end{figure}

To determine the radius of the cluster we created a radial star density profile
centred on the newly measured cluster coordinates. We only included stars in the
profile that possess a photometric uncertainty of less than 0.2\,mag in all
three filters and a quality flag in the Source Extractor photometry of better
than three (see Bertin \& Arnouts \cite{1996A&AS..117..393B} for details). Note
that due to the generally low crowding in the field there is no trend of
photometric uncertainty with distance from the cluster centre, which could
influence the star density profile (shown in Fig.\,\ref{stardens}). 

\begin{figure*}
\centering
\includegraphics[width=14cm]{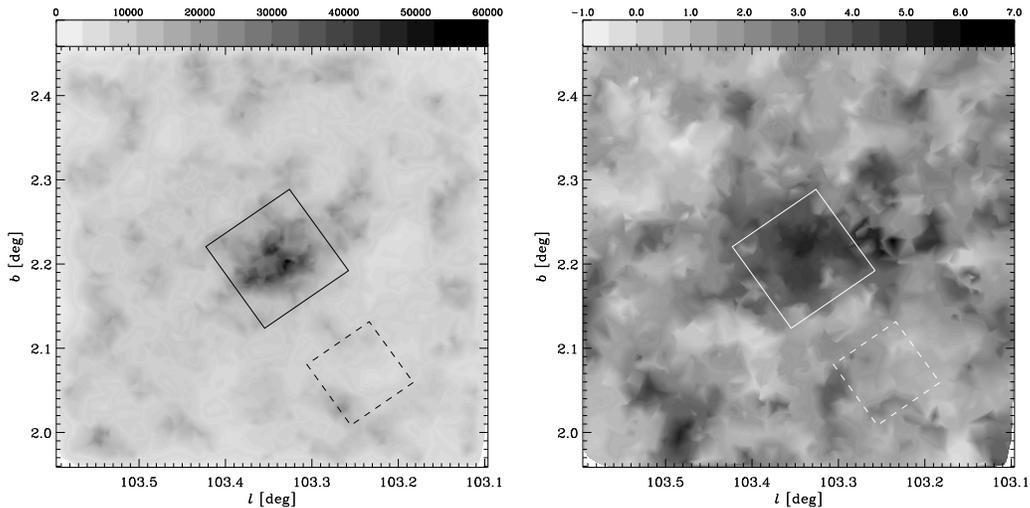}

\caption{\label{2mass_stardens} {\bf Left:} Star Density map of a
0.5\degr\,x\,0.5\degr\ region around FSR\,0358 based on the distance to the
20$^{th}$ nearest neighbour in 2MASS. {\bf Right:} Relative optical extinction
map of the same area based on the median colour excess of the 20 nearest
neighbouring stars. In both panels the solid line square indicates our observed
cluster field and the dashed line square the observed control field i). } 

\end{figure*}

We fit the star density of the cluster using a King-type profile (King
\cite{1962AJ.....67..471K}) and determine a core radius ($r_{core}$) between
40\arcsec and 60\arcsec. The value for $r_{core}$ varies between these two
limits, depending on the coordinates we choose for the centre (since the cluster
centre was so difficult to define we performed several radial density profile
fits). In general, a radius of 60\arcsec\ gives a better fit and we hence choose
this value for $r_{core}$. The tidal radius determined by the fit is of the
order of 400\arcsec, with a rather large uncertainty. This is mostly caused by
the fact that our observations only cover regions up to about 200\arcsec\ from
the cluster centre. Furthermore, the star density around the cluster is variable
by about a factor of 1.5, judging by the 2MASS star density map presented in
Fig.\,\ref{2mass_stardens}. The measured core radius is about a factor of three
smaller than the value listed in Froebrich et al. \cite{2007MNRAS.374..399F}. We
also determined the core radius based on a radial star density profile fit using
solely 2MASS sources in the cluster. This results in a value of
$r_{core}$\,=\,40\arcsec, in agreement with the range of values obtained from
our new data. 

The fitted, background corrected, central star density, when using our new data,
is about 100 stars per square arcminute, while the background star density is
about 60 stars per square arcminute. This shows that the total star density in
the cluster centre is still significantly influenced by the background, but
dominated by cluster stars.

\subsection{2MASS star density and relative extinction maps}
 
To aid our analysis of the small new UKIRT field, we utilise the 2MASS
photometry of a wider area around the cluster. For this purpose we have
determined a star density and a relative extinction map of a
0.5\degr\,x\,0.5\degr\ region around FSR\,0358 in Galactic coordinates. Only
stars with a quality flag of Qflag\,=\,`AAA' are used for these figures. Both
maps are shown in Fig.\,\ref{2mass_stardens}.

The star density map is obtained by measuring the distance to the 20$^{\rm th}$
nearest neighbour of each star. This distance was then converted into the star
density at each position. The resulting map shows a clear enhancement of stellar
density at the position of the cluster. There seems to be a extension of higher
star density towards Galactic North-West. The field around the cluster does not
show any other significant changes in the star density. however, there are small
scale variations on the order of a factor of two.

The relative extinction maps are also calculated by using only the 20 nearest
stars to each position. We determine the median (J-H) and (H-K) colour excesses
of these stars and convert them into extinction values. The effect of using the
median has been extensively discussed e.g. in Froebrich \& del Burgo
\cite{2006MNRAS.369.1901F}. In essence, this type of map shows the extinction
along the line of sight, out to a distance that corresponds to the median
distance of the stars included in the colour excess determination. 

In an area with only field stars, this has the effect that the correct
extinction value is determined, or the cloud is not detected at all. The
distance out to which the extinction can be determined depends on the galactic
position, wavelengths of the observations, the extinction itself and the
completeness limits of the data. Given the simulations in Froebrich \& del Burgo
\cite{2006MNRAS.369.1901F} and using the position of FSR\,0358 we estimate that
the limiting distance for the detection of extinction in the field around the
cluster is between 1\,kpc and 2\,kpc. The exact value is not of interest for
our analysis. Important is only the fact that the limiting distance is much
smaller than the determined distance to the cluster of about 7\,kpc (see
below). 

In the cluster area the situation is different. Here the cluster stars dominate
the numbers, and hence the median distance of the stars corresponds to the
distance of the cluster. Hence in this region the $A_V$ map shows the extinction
along the entire line of sight (about 7\,kpc) to the cluster, and not just the
extinction within 1\,kpc to 2\,kpc.

The resulting map shows that there is a general fluctuation of extinction values
within the field, even on small scales. Within an area of our follow-up
observations there are basically fluctuations of up to 2\,mag $A_V$. Since our
method limits the detectability of variations to 2\,kpc, we expect that the
fluctuations of extinction towards the cluster are at least of the same order of
magnitude. Note that such a scatter in $A_V$ would introduce a scatter in J-K
colours of about $\pm$\,0.25\,mag, more or less exactly what is observed in
colour magnitude diagrams of the area (see below).

At the position of the cluster we find an apparently increased extinction. This
is simply explainable by the much larger distance along the line of sight traced
toward the cluster compared to the surrounding field. This apparent difference
in extinction does not mean that the population of foreground stars changes
systematically between the cluster area and the surrounding field. The measured
value of $A_V$ in the cluster centre is in good agreement with the extinction
towards the cluster determined from isochrone fitting (see below).

\subsection{Field Star Decontamination}

We fit the radial star density using a King profile (King
\cite{1962AJ.....67..471K}). The fit shows that even in the cluster centre, the
population of stars is heavily influenced by fore and background field stars. To
investigate the colour-magnitude (CMD) and colour colour diagrams (CMD) of the
cluster stars we need to statistically decontaminate our sample of stars.

\begin{figure*}
\centering
\includegraphics[angle=-90,width=8.5cm]{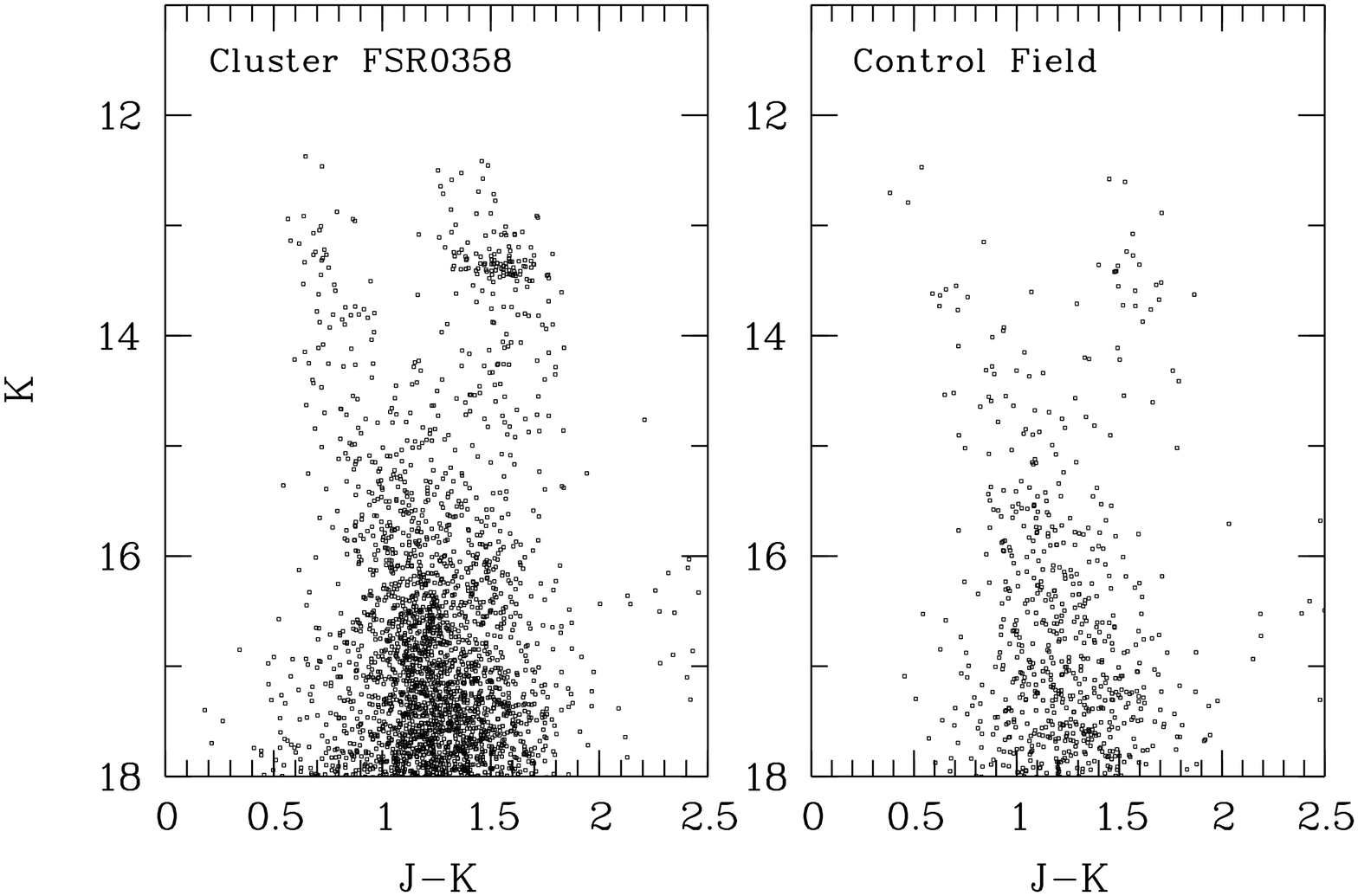} \hfill
\includegraphics[angle=-90,width=8.5cm]{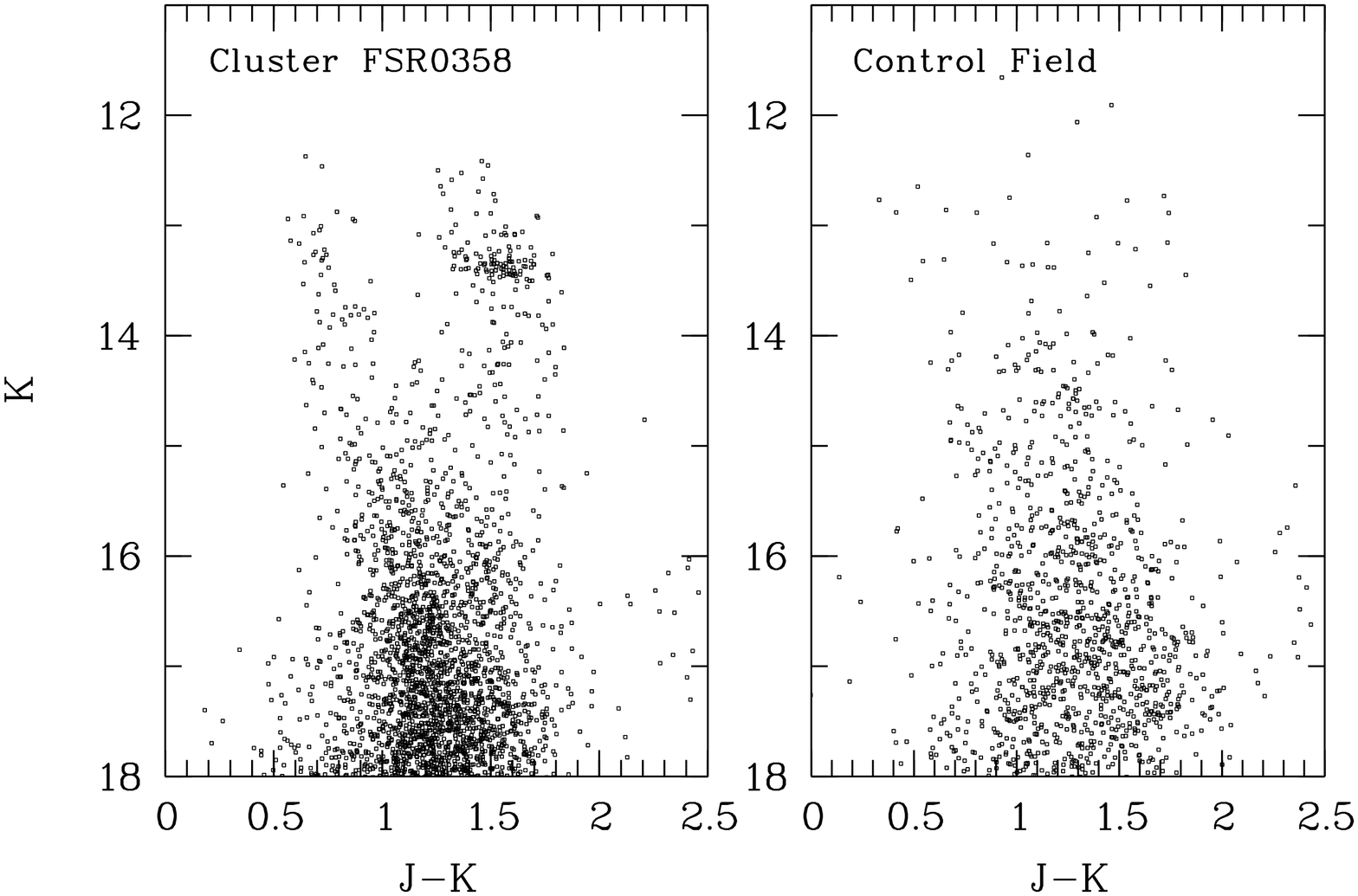}

\caption{\label{twodecon} Calibrated colours and magnitudes of all stars in the
cluster field (1st and 3rd panel) and control fields (2nd and 4th panel). The
2nd panel shows the control ii) field near the cluster, the 4th panel the more
distant control field i) observed in addition to the cluster area.}

\end{figure*}

For this purpose we select all stars within three times the core radius (i.e.
within 3\arcmin\ from the cluster centre) as part of the cluster area. Then we
select two different control fields. (i) all stars in our cluster mosaic that are
more than 3.5\arcmin\ away from the cluster; (ii) all stars in our observed
control field. In Figure\,\ref{twodecon} we show how the distribution of stars
in the CMD of the cluster field and the two control fields compare. 

There are some obvious differences in the distribution of stars in the CMD for
the two control fields. The distribution of stars in control field (i) looks
similar to the cluster field (except for the large number of cluster red
giants). But the control field still contains a few cluster stars, which are
apparent at K\,=\,13.5\,mag and J-K\,=\,1.5\,mag. This is caused by the fact
that we selected stars which are as close as 3.5\arcmin\ to the cluster, while
the tidal radius is more likely 6.5\arcmin. Given the small core radius of the
cluster, however, the number of cluster stars included in control field (i) is
very small (the contribution of cluster stars to the star density at these radii
is less than 1.5\,\%). The control field (ii) shows the same basic distribution
of stars as field (i). However, on close inspection there are some differences.
In particular, the scatter in the J-K colours of the stars is wider in field
(ii). This could be caused by a dust cloud along the line of sight. The fact
that we do not detect a cloud in our extinction map could just mean it is
slightly further away than our limiting distance (see Sect.\,3.3). Some of the
stars in control field (ii) seem to be shifted by up to J-K\,=\,0.4\,mag,
corresponding to an extinction of about $A_V$\,=\,2.3\,mag (Mathis
\cite{1990ARA&A..28...37M}), similar to the scatter in $A_V$ in the extinction
map of our observed control field (see right panel of
Fig.\,\ref{2mass_stardens}). Furthermore, the number of stars per unit area in
control field (ii) is a factor of two lower than in field (i). Given that field
(i) is only 0.1\degr\ closer to the Galactic Plane, extinction has to play an
important role in explaining the differences between the two control fields.

Using the two control fields we decontaminate the cluster area statistically
following the method of Bonatto \& Bica \cite{2007MNRAS.377.1301B}. As cell-size
for the decontamination we choose $\Delta (J)$\,=\,0.6\,mag, $\Delta
(J-H)$\,=\,0.4\,mag, and $\Delta (J-K)$\,=\,0.4\,mag. One realisation of the
decontaminated CMDs and CCDs is shown for both cases in Fig.\,\ref{isochrones}.
We also overplot the 2MASS photometry for bright stars in the cluster area that
were saturated in our images. Because of the differences in the two control
fields, the two decontaminated CMDs appear slightly different. In particular
there are more stars remaining in the decontaminated CMD when using control
field (ii), due to the increased extinction in this field. Note that we
'artificially' increased the field star density for field (ii) before performing
the decontamination. However, the main features (core helium burning objects,
main sequence turn-off and sub-giant branch) are visible in both decontaminated
diagrams.

\subsection{Isochrone Fitting}

To investigate the cluster properties we fit isochrones to the decontaminated
CMDs and CCDs of our photometry, plus the 2MASS photometry of the brighter
cluster stars. We use the isochrones from Girardi et al.
\cite{2002A&A...391..195G} for 2MASS filters. 

As a starting point for our fit we apply the parameters obtained by Kirkpatrick
- a distance of 7.2\,kpc and a reddening of about $A_K$\,=\,0.55\,mag. To fully
characterise the data, we further require the age and metallicity of the cluster
and an extinction law. For the latter we will use the values from Mathis
\cite{1990ARA&A..28...37M} throughout. We then fit isochrones with varying ages
and metallicities simultaneously to both, the decontaminated CMDs and CCDs. As a
fix-point to match the data and the model isochrones we use the red clump stars,
since this is the feature in the data that is the most accurate to determine.

The extinction towards the cluster can be very accurately determined from the
decontaminated CCD, since the isochrone position in this diagram is essentially
independent of the age, distance and metallicity. We find that
$A_K$\,=\,0.6\,$\pm$0.05\,mag fits the data best.

\begin{figure*}
\centering

\includegraphics[angle=-90,width=15.5cm]{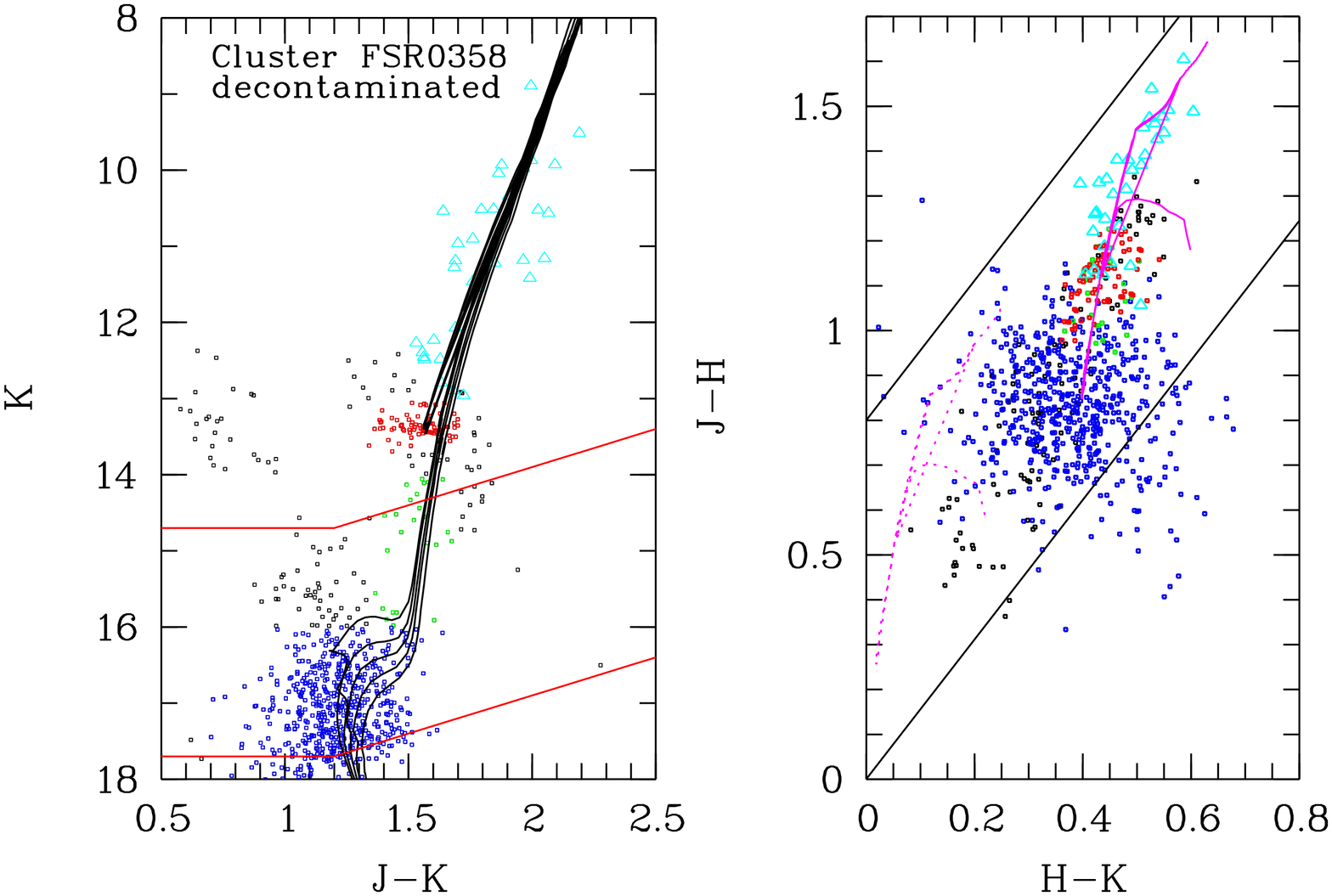} \\
\includegraphics[angle=-90,width=15.5cm]{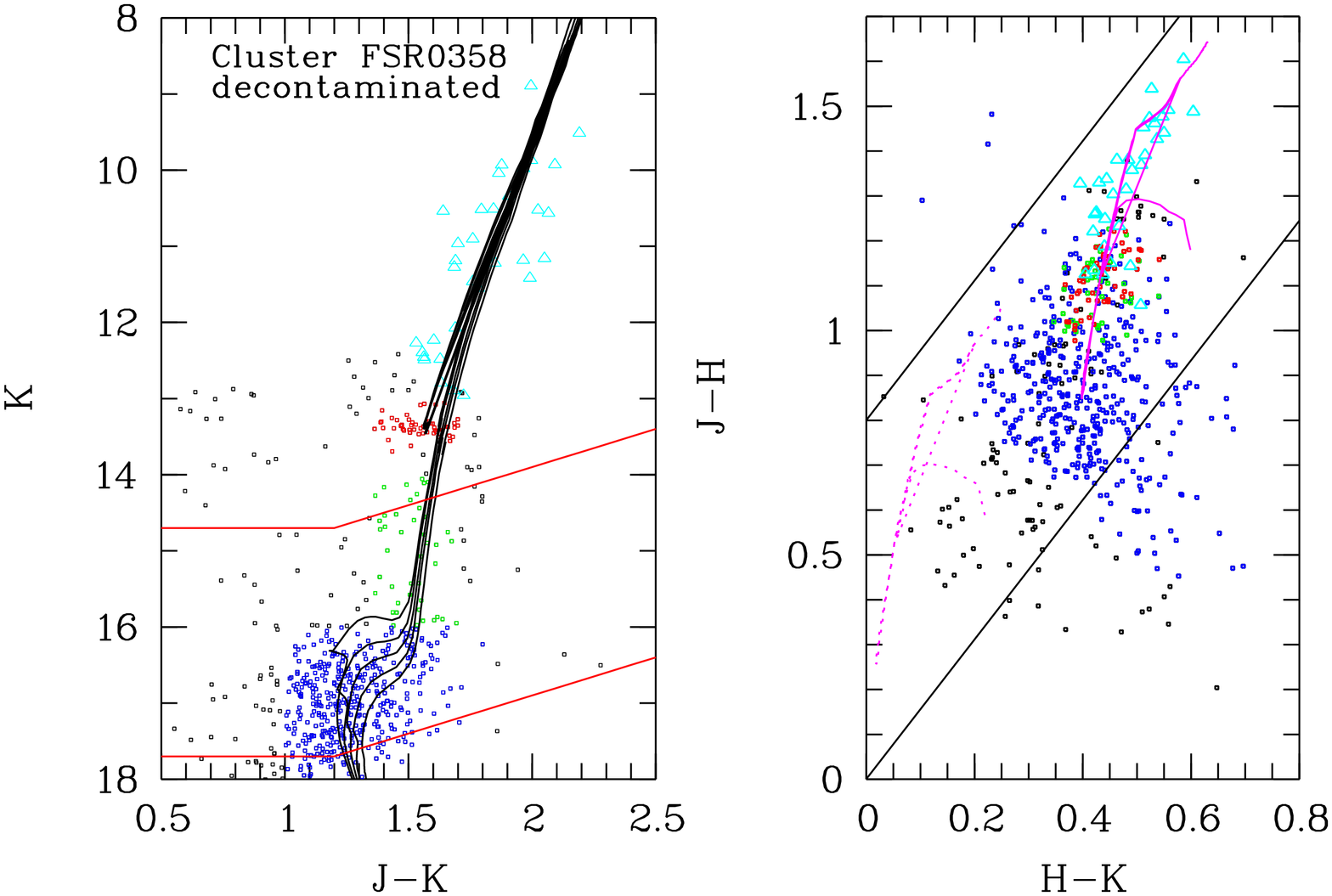}

\caption{\label{isochrones} Decontaminated CMDs and CCDs of the cluster with
overplotted isochrones. The decontamination has been done using the observed
control field i) {\bf (bottom panels)} and the outskirts of the clusters field
itself - control field ii) {\bf (top panels)}. In blue are what we interpret as
the main sequence stars, green are the sub-giants (just leaving the main
sequence), in red are the red-clump Helium burning red giants and as azure
triangles we show the cluster red giants (or AGB stars) detected in 2MASS only
(saturated in our data or brighter than K\,=\,13\,mag). The solid red lines
indicate the completeness limit of our data (bottom) and 2MASS (top) in the
cluster field. The black solid lines in the CMDs are isochrones with Z\,=\,0.008
for a range of ages (log(age)\,=\,9.40, 9.55, 9.70, 9.85, 10.0). The pink solid
line in the CCDs illustrates the best fitting isochrone using age\,=\,5\,Gyr,
Z\,=\,0.008, d\,=\,7.2\,kpc, $A_K$\,=\,0.6\,mag.}

\end{figure*}

We then fit isochrones with varying ages (log(age)\,=\,9.40, 9.55, 9.70, 9.85,
10.0) and different metallicities (Z\,=\,0.004, 0.008, 0.019) to the data. Each
isochrone requires a slightly different extinction and distance of the cluster
to fit the red clump stars. As an example we show in the left panels in
Fig.\,\ref{isochrones} all isochrones for Z\,=\,0.008.
The various isochrones are then used to determine the best fitting parameters
and their uncertainties. 

At first we constrain the metallicity. It can in principle be determined using
the slope of the RGB stars in the J-K vs. K diagram and using e.g. Fig.\,8 in
Valenti et al. \cite{2004A&A...419..139V}. Applying this method we find
[M/H]\,=\,-0.4\,$\pm$\,0.2\,dex or Z\,=\,0.005 to Z\,=\,0.012. We can also look
at the required extinction values for the isochrones. For Z\,=\,0.004 a range of
$A_K$\,=\,0.63 to 0.64\,mag (depending on the age) is required. In the case of
solar metallicity (Z\,=\,0.019) we require a range of $A_K$\,=\,0.54 to
0.56\,mag to obtain a fit in the CMD. Thus, for these cases the range of
required $A_K$ values is on the borders of our uncertainty limit for the
extinction. In particular the solar metallicity does not fit the data, and for
Z\,=\,0.004 only a marginal agreement with the data can be found. For
Z\,=\,0.008, however, the required range of extinction values to fit the
isochrones is $A_K$\,=\,0.60 to 0.61\,mag. This is in good agreement with our
$A_K$ estimate using the CCD. Thus, the metallicity of the cluster can be
constrained to Z\,=\,0.008\,$\pm$\,0.004.

To establish the correct distance to the cluster we need to constrain the age.
The usual procedure is to use the main sequence turn-off position. Though the
turn-off is detected in our data its exact position in the CMD is difficult to
determine. This is mostly due to the scatter of the photometry due to variable
extinction in the field. However, in the decontaminated CMDs we can identify a
number of sub-giant branch stars, which we can use as a further aid to ascertain
the best fitting isochrone. Of the five isochrones we fit to the data, the one
with the lowest age (log(age)\,=\,9.40) cannot explain the observed sub-giant
branch. This is indeed also the case for all other metallicities. The remaining
isochrones fit the data well. However, the 10\,Gyr isochrone clearly has a
turn-off that is at too faint magnitudes. Hence, the age of the cluster can be
constrained to between 3.5\,Gyr and 7.0\,Gyr with the most likely value of
5\,Gyr. Please note that the distribution of age uncertainties might be even
more asymmetric than suggested by the above range. This is caused by the fact
that the main sequence turn off is near the photometric limit. Hence, the
brightnesses of the cluster stars might be systematically increased due to noise
and blends, which also might change the colours. Thus, the 5\,Gyr are likely an
upper limit.

If we use the 5\,Gyr isochrone, Z\,=\,0.008, and an extinction of
$A_K$\,=\,0.6\,mag, the distance of the cluster is 7.2\,kpc. Considering the
ranges in age, Z, and $A_K$ that we established, a range in distances from
6.8\,kpc to 7.7\,kpc is found. Hence we can constrain the distance to FSR\,0358
to 7.2\,$\pm$\,0.5\,kpc. The pink lines in the CCDs in Fig.\,\ref{isochrones}
are drawn using the best fitting parameters. Figure\,\ref{isochrones} also
illustrates that these results do not significantly depend on the choise of the
control field for field star decontamination.

We obtain the same distance as given by Kirkpatrick. Our extinction is slightly
higher, which is caused by our finding that the metallicity is sub-solar, in
contrast to the assumption made in the original analysis. Furthermore, our new
deeper data allowed us to stronger constrain the age of the cluster.

 %
 %
 %
 %

\subsection{Deduced Cluster Properties}

Given our isochrone and radial star density fit, we conclude that FSR\,0358 is
an old stellar cluster. Judging by the determined cluster parameters and its
optical appearance, this object is similar to FSR\,0190 (Froebrich et al. 
\cite{2008MNRAS.383L..45F}). Both clusters have ages above several Gyr and
sub-solar metallicities. The parameters put the cluster at a distance of
11.3\,kpc from the Galactic Centre and 275\,pc above the Galactic Plane, typical
values for old open clusters. Note that FSR\,0190 was found to be 10.5\,kpc away
from the Galactic Centre and 170\,pc above the plane (Froebrich et al.
\cite{2008MNRAS.383L..45F}). These galactocentric distances are derived assuming
a distance of the Sun to the Galactic Centre of 7.2\,kpc (Bica et al.
\cite{2006A&A...450..105B}).

Using the determined distance and radial star density fit, the core radius of
the cluster is 2\,pc. We can estimate its original mass following Salaris \&
Girardi \cite{2002MNRAS.337..332S}. The current number of about 75 core helium
burning objects indicates that FSR\,0358 started off as an about
10$^5$\,M$_\odot$ cluster. We determine the clusters current absolute magnitude
by adding the K-band magnitudes of all detected giant stars (our data plus
2MASS) and correct for the determined distance and extinction. We obtain
M$_K$\,=\,-8.1\,mag, or M$_V$\,=\,-6.2\,mag assuming $V-K$\,=\,1.9\,mag
(Leitherer et al. \cite{1999ApJS..123....3L}). This puts it at the bright end of
the open cluster luminosity function (van den Bergh \& Lafontaine \cite{1984AJ.....89.1822V}). We summarise all the determined
cluster parameters in Table\,\ref{summary}. 

\begin{table}
\caption{\label{summary} Summary table of the determined properties of FSR\,0358.}
\begin{tabular}{ll|rr}
Property & value & Property & value \\ \hline
RA (J2000) & 22:10:14.1 & DEC (J2000) & $+$58:48:02 \\
$l$ & 103.34094\degr & $b$ & $+$2.20214\degr \\
age & 5\,$\pm$\,2\,Gyr & Z & 0.008\,$\pm$\,0.004 \\
d$_\odot$ & 7.2\,$\pm$\,0.5\,kpc & A$_K$ & 0.6\,$\pm$\,0.05\,mag \\
d$_{\rm GC}$ & 11.3\,$\pm$\,0.8\,kpc & z & $+$275\,$\pm$\,20\,pc \\
$r_{core}$ & 60\arcsec\ & $r_{core}$ & 2.1\,$\pm$\,0.2\,pc \\ 
M & 10$^5$\,M$_\odot$ & M$_V$ & -6.2\,mag \\
\end{tabular}
\end{table}

\section{Conclusions}
\label{discussion}

Using an expectation-maximisation method, we have identified FSR\,0358 as the
most likely star cluster among the entire list of clusters in Froebrich et al.
\cite{2007MNRAS.374..399F}. It was originally discovered by Kirkpatrick and
classified by this author as an old open star cluster. Our analysis of new UKIRT
follow-up observations in combination with 2MASS data essentially confirms this
conclusion, puts it on a more solid ground and allows us to refine some cluster
parameters. Based upon the large population of red giant stars, the
well-populated red clump in combination with the subgiant branch and the main
sequence turn-off we derive a narrower age interval of $5\pm2$\,Gyr and a lower
metallicity of $Z=0.008\pm0.004$ compared with Kirkpatrick. Our distance of
7.2\,kpc is in agreement with the original value. This corresponds to a large
distance from the Galactic centre, typical for old open clusters. Note that
these results do not significantly depend on the choise of the control field for
the field star decontamination. The object is notably similar to FSR\,0190,
another old open cluster detected amongst the FSR candidates (Froebrich et al.
\cite{2008MNRAS.383L..45F}). 

The confirmation of FSR\,038 (Kirkpatrick\,1) as another well populated,
distant, old open cluster shows that the FSR sample is useful for characterising
the entire population of clusters within the Milky Way. It not just allows us to
detect embedded young clusters, but also enables us to improve our statistics on
the oldest clusters in the Galaxy, so far hidden behind dust clouds near the
Galactic Plane. There are, for example, only 12 clusters with ages above 5\,Gyr
listed in the WEBDA database\footnote{\tt http://www.univie.ac.at/webda/}. So
far there have been six clusters identified in the FSR list with an age of about
5\,Gyr or older, significantly enhancing the known sample of these old clusters
(FSR\,0358 - this paper; FSR\,1735 - Froebrich et al. \cite{2007MNRAS.377L..54F,
2008MNRAS.390.1598F}; FSR\,0070, 1737 - Bica et a. \cite{2008MNRAS.385..349B};
FSR\,0584 - Bica et al. \cite{2007A&A...472..483B}; FSR\,0190 - Froebrich et al.
\cite{2008MNRAS.383L..45F}; and possibly FSR\,1767 - Bonatto et al.
\cite{2007MNRAS.381L..45B} but see Froebrich et al. \cite{2008MNRAS.390.1598F}).

\section*{acknowledgments}

We would like to thank the referee S.\,Ortolani for pointing us to the
previous works by J.D.\,Kirkpatrick on this cluster and for further helpful
comments to improve the analysis of our data. SS is supported by the Deutsche
Forschungsgemeinschaft (DFG) through grant SCHM 2490/1-1. The United Kingdom
Infrared Telescope is operated by the Joint Astronomy Centre on behalf of the
Science and Technology Facilities Council of the U.K. This publication makes use
of data products from the Two Micron All Sky  Survey, which is a joint project
of the University of Massachusetts and the Infrared Processing and Analysis
Center/California Institute  of Technology, funded by the National Aeronautics
and Space  Administration and the National Science Foundation.

\label{lastpage}

\end{document}